\input harvmac
\hfuzz=100pt
 
\overfullrule=0pt
\parindent=0pt


\def\G(#1){\Gamma(#1)}

\def\C|#1{{\cal #1}}
\def\(#1#2){(\zeta_#1\cdot\zeta_#2)}
\def\lr{\lref}


\def\xxx#1 {{hep-th/#1}}
\def\lr { \lref}
\def\npb#1(#2)#3 { Nucl. Phys. {\bf B#1} (#2) #3 }
\def\rep#1(#2)#3 { Phys. Rept.{\bf #1} (#2) #3 }
\def\plb#1(#2)#3{Phys. Lett. {\bf #1B} (#2) #3}
\def\prl#1(#2)#3{Phys. Rev. Lett.{\bf #1} (#2) #3}
\def\physrev#1(#2)#3{Phys. Rev. {\bf D#1} (#2) #3}
\def\ap#1(#2)#3{Ann. Phys. {\bf #1} (#2) #3}
\def\rmp#1(#2)#3{Rev. Mod. Phys. {\bf #1} (#2) #3}
\def\cmp#1(#2)#3{Comm. Math. Phys. {\bf #1} (#2) #3}
\def\mpl#1(#2)#3{Mod. Phys. Lett. {\bf #1} (#2) #3}
\def\ijmp#1(#2)#3{Int. J. Mod. Phys. {\bf A#1} (#2) #3}

\def\lam16{\lambda^{16}}

\parindent 25pt
\overfullrule=0pt
\tolerance=10000

\sequentialequations

\noblackbox
\baselineskip 14pt plus 2pt minus 2pt

\Title{\vbox{\baselineskip12pt
\hbox{hep-th/9712187}
\hbox{CPTH-S592.1297}
\hbox{DAMTP-97-146}
}}
{\vbox{\centerline{A Classical Manifestation  of the }
\vskip 0.2cm
\centerline{ Pauli Exclusion Principle}}}
\bigskip
\centerline{ Constantin P. Bachas }
\medskip
\centerline{  Centre de Physique Th\'eorique,  Ecole
Polytechnique,}
\centerline{ 91128 Palaiseau, France}
\centerline{\it bachas@pth.polytechnique.fr}
\smallskip
\centerline{and}
\smallskip
{\centerline{Michael B.  Green}}
\medskip
\centerline{DAMTP, Silver Street, Cambridge CB3 9EW, UK}
\centerline{\it  M.B.Green@damtp.cam.ac.uk}
 \bigskip

\centerline{\bf Abstract}

The occupied and unoccupied  fermionic BPS quantum states of a type-IIA
string stretched between a D6-brane and an orthogonal  D2-brane 
are described in M-theory by  two particular  holomorphic curves embedded
in a Kaluza--Klein monopole. 
The absence of multiply-occupied fermionic states ---
 the Pauli exclusion principle ---  is manifested in M-theory
by the absence of any
 other holomorphic curves satisfying the
necessary boundary conditions.
Stable, non-BPS states with multiple strings joining
 the D6-brane and D2-brane are described M-theoretically
 by non-holomorphic curves.

\Date{December 1997}
 \vfill\eject
\lr\ohta{T. Nakatsu, K. Ohta, T. Yokono and
Y. Yoshida,
 {\it A proof of brane creation via M-theory}, hep-th/9711117.}
\lr\yoshida{Y. Yoshida, {\it Geometrical Analysis of Brane Creation
 via M-Theory}, hep-th/9711177.}
\lr\hw{ A. Hanany and E. Witten,
 {\it Type-IIB Superstrings, BPS
Monopoles and Three-Dimensional Gauge Dynamics},  hep-th/9611230;
 Nucl.Phys. {\bf B492} (1997) 152.}
\lr\bdg{C.P.  Bachas, M.R.  Douglas and M.B.  Green, {\it
Anomalous creation of branes},  hep-th/9705074;
JHEP07 (1997) 2.}
\lr\dfk{U. Danielsson, G. Ferretti and  I.R. Klebanov, {\it
Creation of Fundamental Strings by Crossing D-branes},  hep-th/9705084;
Phys. Rev. Lett. {\bf 79} (1997) 1984.}
\lr\bgl{ O.  Bergman, M.  Gaberdiel and G.  Lifschytz, {\it Branes,
Orientifolds and the Creation of Elementary Strings}, hep-th/9705130.}
\lr\bgs{C.P. Bachas, M.B. Green and  A. Schwimmer,
{\it (8,0) Quantum mechanics and symmetry enhancement
 in type I' superstrings},
 hep-th/9712086.}
\lr\ov{H. Ooguri and  C. Vafa, {\it Geometry of N=1 Dualities in
 Four Dimensions}, 
 hep-th/9702180;
 Nucl.Phys. {\bf B500} (1997) 62.} 
\lr\alwis{S. P. de Alwis, {\it A note on brane creation}, 
 hep-th/9706142.}
\lr\ho{P-M.  Ho and Y-S. Wu, {\it Brane Creation in M(atrix)
Theory}, hep-th/9708137.}
\lr\osz{N. Ohta, T. Shimizu and  J-G. Zhou, {\it
Creation of Fundamental String in M(atrix) Theory}, 
hep-th/9710218.}


\noblackbox
\baselineskip 14pt plus 2pt minus 2pt

A striking aspect of duality symmetries is that they exchange 
classical and quantum properties of the system under study.
For instance T-duality exchanges the 
momentum,  $n{\hbar}/R$,  of a 
closed  string with the energy stored in  winding,  $2\pi n  T_F R$, where
$T_F$ is the fundamental
string tension. The quantization of winding is a classical
geometric property of large smooth strings while the quantization of 
momentum  is a consequence of the single-valuedness
of the quantum mechanical wave function.
 T-duality exchanges these  classical and quantum  features.
Such relationships between classical and quantum phenomena
should arise naturally in the  M-theory description of string theory. 

 One principle that appears to be essentially quantum mechanical
 is the   Pauli exclusion principle,
which  forbids two fermions from occupying the same quantum state. 
In this note we  will point out how this can arise classically in 
 an M-theory context. We will consider an M-theory membrane in the background
of a Kaluza-Klein monopole, which is the M-theory description of a
 six-brane.  The membrane is placed so as to be asymptotically  transverse
 to the  six-brane. In the type IIA limit
this configuration is described by a D6-brane and  an orthogonal D2-brane
 spaced a finite distance apart. The  configuration has  a total of eight
(DN)  directions that are  in one brane and transverse to the other 
and it  preserves one quarter of the 32 components of 
 the M-theory supersymmetry.  This setup is very special since
a fundamental string  stretching between the two D-branes has a unique,
localized BPS  state which is necessarily a fermion. As a result there
are  only two independent BPS quantum states
in the second-quantized theory  ---
 the fermionic string is either present or absent.
  In the  M-theory description of this system 
each of these configurations 
is  described by a  single smooth membrane which must be
 holomorphically embedded  in the Kaluza--Klein monopole 
 in order for the BPS condition to be satisfied.   In this
 particular context the  Pauli
exclusion principle will turn out to be the statement  that
there are only two such 
  holomorphic curves that satisfy the boundary conditions.

Solitonic p-branes that are fermions usually arise from the quantization
of  zero modes  and are part of a supermultiplet which also includes
some bosonic states. Furthermore they are usually free to move in one
or more spatial directions, and  therefore have an  infinite number of allowed
quantum states. 
In such circumstances the   Pauli exclusion principle would not
 normally have a  direct geometric meaning. The configuration under
 discussion here  is exceptional
 because the membrane dual to the
fundamental string is necessarily fermionic and is localized.

  The same statement is true for any system related to
 this by S or T dualities.
In fact, a system of  this type  was
used   to understand two of the subtle rules of gauge field theory engineering
with branes \hw --- the anomalous creation of branes
   and the so-called s-rule which
forbids certain supersymmetric configurations of branes from existing.
  Brane creation  can be  explained in terms of the anomalous inflow
 of charge \bdg\  (see also \dfk\bgl\ho\osz\ for related arguments)  and 
 in geometric terms  
\ov\alwis\ohta\yoshida. The s-rule can be argued   \bgs\ to be a consequence
 of the Pauli exclusion principle. This note  builds on the
 interrelationship between these observations.

   The type IIA configuration of interest has a D6-brane along the
directions $(x_0,x_1, x_2,x_3,x_4,x_5,x_6)$, and a D2-brane along
 $(x_0, x_7, x_8)$. The two are separated in the $x_9$ direction.  From 
the M-theory point of view the D6-brane is a Kaluza--Klein monopole 
described by the metric,
\eqn\taub{
ds^2 =\;  -(dx_0)^2\;  +\;  \sum_{i=1}^6 (dx_i)^2\;  +\;   V\;
 d{\vec r}\cdot 
d{\vec r} + V^{-1}
( dx_{10} + {\vec A}\cdot d{\vec r})^2\ , }
where 
\eqn\NUT{ V = 1 + {1\over 2\vert {\vec r} \vert} \ .}
Here 
$x_{10}= x_{10}+2\pi$ is the angular 
coordinate, 
  ${\vec r}\equiv  (x_7, x_8, x_9)$, and ${\vec A}$
is the monopole field which satisfies the equation ${\vec \nabla} \times
{\vec A} = {\vec\nabla} V$. Length scales are measured in units of
the radius of the circle on which M-theory is compactified,
and  which has been set equal to one. The monopole
is located at ${\vec r} =0$.

The Taub-NUT space has three covariantly-constant complex structures.
One particular choice of  holomorphic coordinates  is \ohta\yoshida
\eqn\coord{
v = x_7+ix_8\ , }
and
\eqn\coordi{
w  = e^{-(x_9+ix_{10})} \left( - x_9 + \sqrt{ x_9^2 + \vert v^2 \vert
 } \right)^{1/2} .}
In these coordinates the metric takes the K\"ahler form
\eqn\Kahler{
ds^2 =\;  -(dx_0)^2\;  +\;  \sum_{i=1}^6 (dx_i)^2\;  +\;   V\;
dv d{\bar v}\; + V^{-1} \left\vert {dw\over w} - f  dv \right\vert^2\ 
,}
where
\eqn\delt{
f = {x_9 +\vert {\vec r}\vert \over
2 \vert v \vert \vert {\vec r}\vert} \ . }
The  metric  is smooth everywhere except for a
 coordinate singularity along the positive $x_9$-axis.

Any holomorphic curve $w(v)$ describes a BPS configuration of a membrane
in this metric. Two simple curves are 
\eqn\curves{{\rm (a)}\qquad  w = e^{-b}\ ,\qquad {\rm (b)} \qquad 
w = e^{-b} v \ ,}
with $b$ an  arbitrary real constant. The first curve describes a 
membrane at fixed $x_{10}$, while in the second $x_{10}$ winds by
$2\pi$ as $v$ circles  clockwise around the origin.
The radial profiles
of these two curves are mirror reflections of each other,
\eqn\static{ 
\vert v\vert =  e^{\pm (x_9-b)} \sqrt{\pm  2x_9 +  e^{\pm 2(x_9-b)}} \ ,  }
where the plus sign corresponds to the first curve, and the minus
sign to the second. 

The type IIA limit is reached by taking $\vert x_9\vert , 
 \vert v\vert$ and $\vert b \vert $
to infinity, while keeping their ratios finite. This corresponds to
blowing up all scales compared to the radius of the compactification
circle. For  $\vert v\vert$ much larger than all other scales,
 $x_9$ approaches
$b$ asymptotically, modulo logarithmically-small corrections. Both curves
therefore describe   a planar D2-brane located at $x_9\simeq b $. 
However, the  two curves differ  drastically in  the region 
$v \sim 0$, where the membrane intersects the $x_9$ axis.
Their behaviour there  depends on the sign of $b$.  For example,
 if  $b$
is  positive then   $x_9 \simeq 0$ for   curve  (a),
while $x_9 \simeq b$ for  curve (b), in this region.
 Therefore, when  $b>0$  the first curve  describes 
a D2-brane located to the right of the D6-brane
and attached to it  by a stretched string.
  On the other hand the second curve describes
a D2-brane on  the right of the transverse D6-brane  without
any strings attached to it.
When  $b<0$ the D2-brane is to the left of the D6-brane and 
the roles of two curves are interchanged.  Now it is curve (b) that
describes the state with a string joining the branes while
curve (a) describes the disconnected branes.

This geometry of one of the two 
 curves (curve (b)) was described in detail in \ohta \yoshida,
 where it was related to  the M-theory description of
 the phenomenon of anomalous brane creation.   However, both  curves
 are needed to make the connection to brane creation in the IIA theory
 complete.  Following the   anomaly inflow discussion in \bdg\
 the process
 of brane creation should be viewed as a phenomenon in which second
 quantization  plays an essential r\^ole. 
In the region $b<0$, and in the type IIA limit, a first quantized string
stretching between the D-branes has a negative-energy fermionic level.
Curve (a) describes the   state in which this level
is occupied, while curve (b) describes the state in which this level
is empty.  In a second-quantized formalism 
curve (a) thus  describes  the vacuum state of the Dirac sea,  in which all
 negative energy levels are occupied and positive levels empty, while
curve (b) describes a physical hole excitation. 
As $b$ is varied adiabatically, the first-quantized energy levels
shift uniformly, 
 just as they do in compact $(1+1)$-dimensional QED when an electric field
is slowly turned on. In the region $b>0$ the BPS energy level has
crossed to positive energy, so what was previously the ground state of
the Dirac sea (described by curve (a))
 is reinterpreted as the state with a single physical fermion 
 with energy proportional to $b$, indicating the presence of the string. 
Likewise, the hole state (curve (b)) is reinterpreted as the new 
second-quantized ground state.
The presence of the two curves that are related by mirror reflection
 is essential for this interpretation,  and the
 interchange symmetry is a manifestation of the CPT symmetry
 of the second-quantized IIA description.

Type IIA string theory in the background of 
 a D6-brane and a transverse D2-brane
has no other finite-energy BPS states, beyond those described by the above
two curves.
To see that no other holomorphic curves are possible in M-theory  consider 
a more general curve $F(w,v)=0$. The degree of $F$ considered
as a polynomial in $w$ determines the number of 
D2-branes oriented in the $(x_7, x_8)$ directions in the string
 theory limit.  This is the number of solutions 
 of the equation for fixed $v$. Since we want to describe a 
single D2-brane we will take the degree to be one,
 so that the curve must have the form 
\eqn\curve{ w = C(v)\ .}
However,  we also require that  the  membrane is  at finite $x_9$ for all
finite values of $v$.   Therefore, since $w$ only diverges
 at $x_9\to -\infty$, the
 meromorphic function $C(v)$ cannot have  poles in the complex $v$-plane.
Such poles would  correspond to semi-infinite strings attached to
the D2-brane in the type IIA limit and stretching to the left.
 
Consider next the zeros of $w$, which  can appear either at
$x_9\to \infty$ and any $v$, or at $v=0$ and any $x_9$. The first
correspond to semi-infinite strings
stretching to the right, in the type IIA limit,   and are
not allowed. Therefore, the most general form for $C(v)$ is $C = a v^n$.
But the form \coordi\ of the $w$-coordinate shows that $\vert w \vert$
 vanishes 
{\it at most} as fast as $\vert v \vert $ at the origin. The only allowed
values are therefore
 $n=0,1$ . This proves the  claim that there are only two holomorphic
curves, with  no infinite strings attached, and they correspond precisely
to the two quantum states discussed above.

Of course, it is very easy to find surfaces
that are not holomorphically embedded and, in the IIA limit,
describe two or more  fundamental strings joining a D6-brane and a D2-brane. 
   However, these are not BPS states and break supersymmetry.
  The ground state of such a configuration would define a stable
 non-supersymmetric vacuum state of a quantum field theory
 living in the membrane. In the IIA limit the mass gap
between the lowest BPS string state and excited string states is
of  order  $\hbar (2\pi \alpha')^{-1/2}$. It would be 
 interesting
to determine  the energy of such an    M-theory configuration
 directly, but   we have not investigated this issue.
The system studied  here transforms, 
through a chain of  dualities,  to the system considered in \hw\ consisting
 of a  D5-brane and a NS 5-brane.   These  may be joined by a single
 BPS D3-brane  but  states with   multiple D3-branes are not supersymmetric.
 This is the s-rule that is in accord with the absence of supersymmetric
 vacua for three-dimensional $U(k)$ supersymmetric Yang--Mills theory
with Fayet-Iliopoulos term
 for $k>1$.     It might be of interest to see whether this mechanism
 can be generalized to apply to 
supersymmetry breaking in more realistic contexts.

\vskip 0.5cm

 {\it Aknowledgements}:
 We thank S. Theisen for a discussion. This work was partially supported by
the EEC grant  TMR-ERBFMRXCT96-0090.

\listrefs

 \bye